\documentclass[cits]{PoS}

\usepackage{subfigure}
\usepackage{booktabs,amsmath}

\newcommand{\rmO}{\mathrm{O}}
\newcommand{\bg}{b_{\rm g}}
\newcommand{\mq}{m_{\rm q}}
\newcommand{\tr}{{\rm tr}}
\newcommand{\Nf}{N_{\rm f}}
\newcommand{\nf}{N_{\rm f}}
\newcommand{\fK}{f_{\rm K}}
\newcommand{\texp}{\tau_{\rm exp}}
\newcommand{\tint}{\tau_{\rm int}}
\newcommand{\tnotc}{t_0^{\rm chiral}}
\newcommand{\wnotc}{w_0^{\rm chiral}}
\newcommand{\tnotp}{t_0^{\rm phys}}
\newcommand{\wnotp}{w_0^{\rm phys}}

\title{On the $\Nf$-dependence of gluonic observables}

\newcommand{\preprintline}{\vspace{3cm}\newline
\rightline{\parbox{2.9cm}{\large\tt DESY 13-216}}
}

\ShortTitle{On the $\Nf$-dependence of gluonic observables}

\author{\speaker{Mattia Bruno} and Rainer Sommer\\
        NIC @ DESY. Platanenallee 6, 15738 Zeuthen, Germany\\
        E-mail: \email{mattia.bruno@desy.de}}
\author{for the ALPHA collaboration}

\abstract{We compute $t_0$, $w_0$ and the topological susceptibility,
defined at finite gradient flow time for two-flavour QCD. 
The use of three lattice spacings and pion masses between 192 and 500 MeV together with a careful error analysis allow to approach the continuum limit of the two-flavour theory despite significant auto-correlations. A comparison 
to $\nf=0$ results shows the size of sea quark effects in $t_0^2\chi$, with
$\chi$ the topological susceptibility, and low energy observables such as
$t_0/w_0^2$ and $t_0/r_0^2$.
\preprintline}

\FullConference{31st International Symposium on Lattice Field Theory - LATTICE 2013\\
		July 29 - August 3, 2013\\
		Mainz, Germany}

\begin{document}

\section{Introduction}

In these Proceedings we investigate the effects of
dynamical quarks on gluonic observables. In the high-energy sector 
the dominant effect of dynamical quarks is well understood in terms of
the running of the coupling. 
To probe the low-energy sector we use the recently introduced 
gradient flow, which allows to define new renormalisation-group-invariant observables~\cite{wflow_luscher}. Here we
concentrate on $t_0$ and $w_0$ as well as 
the topological susceptibility. We compute them with 
$\Nf=2$ light dynamical quarks using the CLS ensembles, based on 
$\rmO(a)$-improved Wilson fermions (see \cite{lat13:stefano}
for the set of ensembles), and we compare
with $\Nf=0$ and $\Nf>2$ results from the literature.
Clearly, the topological susceptibility
 is of particular interest since
chiral perturbation theory predicts a suppression 
proportional to the quark mass for light sea quarks. 
However, critical slowing down of the topological modes
makes the susceptibility particularly difficult to study~\cite{virotta_errors}. Taken together this means that the susceptibility is an important 
indicator for the 
correctness of the simulations. 

The gradient flow is defined by the following equation
\cite{wflow_luscher}:
\begin{align}
\partial_t B_\mu(x,t) &= D_\nu G_{\nu \mu} (x,t)\,, \quad B_\mu(x,0)=A_\mu(x) \,,
\\
G_{\mu \nu} &= \partial_\mu B_\nu - \partial_\nu B_\mu + [ B_\mu ,B_\nu ] \,, \quad D_\mu = \partial_\mu + [B_\mu, \cdot ] \,,
\end{align}
where $B_\mu(x,t)$ is the gauge field at positive flow time $t$ (which has dimension length squared). The energy density 
$\langle E(t) \rangle = -\frac{1}{2} \langle {\rm tr} \{ G_{\mu \nu} G_{\mu \nu} \}\rangle$
has been used to define a scale $t_0$ via 
$\left. t^2 \langle E(t) \rangle \right|_{t=t_0} = 0.3 $.
In general, the correlation functions of the  smooth field $B_\mu(x,t)$ 
do not need renormalisation at any separation in space-time. Therefore
$q(x,t)=-\frac{1}{32 \pi^2} \epsilon^{\mu \nu \rho \sigma} \mathrm{tr} \{ G_{\mu \nu}(x,t) G_{\rho \sigma}(x,t) \}$ can be used directly to define the topological charge $Q(t)=a^4\sum_x q(x,t)$. 
We evaluate it at $t=t_0$ using the clover-type 
(``symmetric'', cf.~\cite{wflow_luscher}) discretisation of $G_{\mu \nu}(x,t)$.

\section{Topology and auto-correlations}

\begin{figure}[t!] 
\centering
\includegraphics[trim= 0 2pt 0 0, width=.47\textwidth]{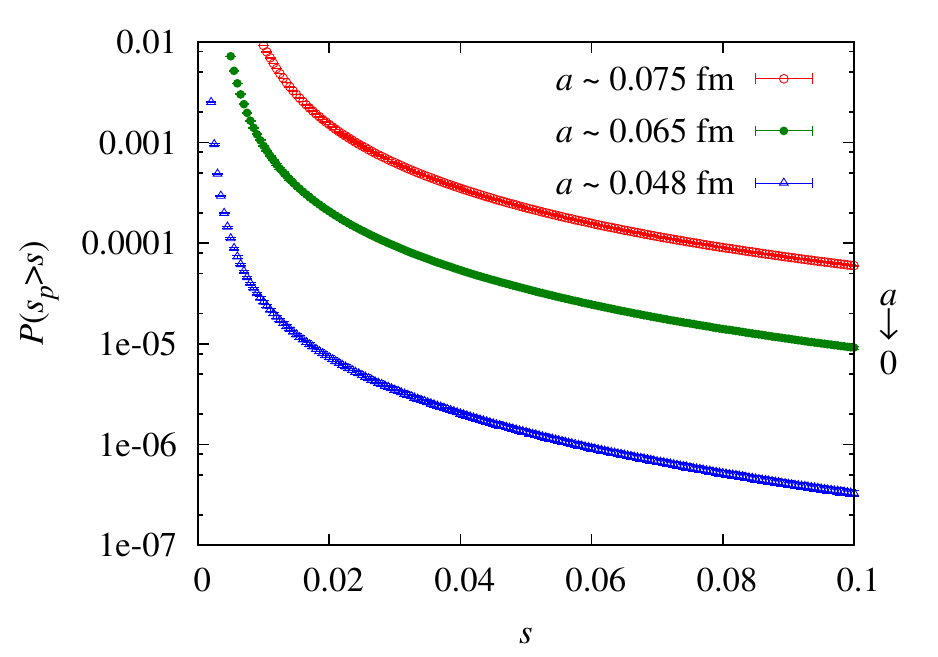}
\includegraphics[trim= 0 2pt 0 0, width=.47\textwidth]{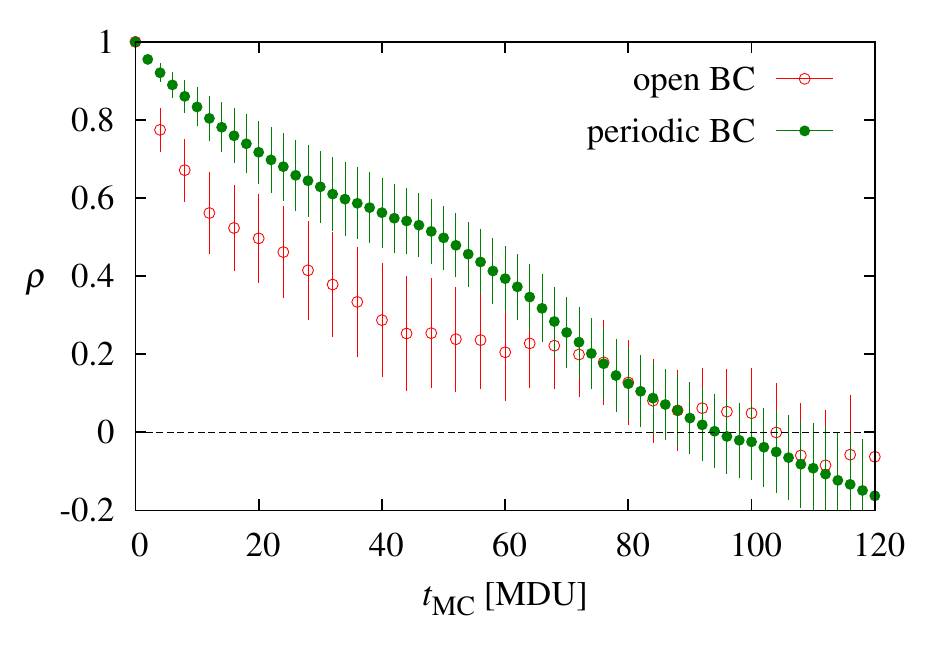}
\caption{{\it Left}: 
probability of $s_p(t)={\rm Re}\,{\rm tr} ( 1-V_t(p) )$ (where $V_t$ is the plaquette loop) to be bigger than a certain value $s$
at fixed $r_0 m_\pi\approx 0.6$. 
Note that for $s_p < 0.067$ the space of lattice fields consists of disconnected sectors \cite{Luscher:1981zq}. {\it Right}: normalised auto-correlation function of $t_0$ with periodic and open BC with $a=0.075$~fm, $m_\pi \approx 280$~MeV and roughly 1000 MDU. }
\label{fig1}
\end{figure}

Following \cite{wflow_luscher} we check how strong the 
separation of topological sectors is realised with our lattice 
action. 
In principle the suppression of regions in configuration space 
between the charge sectors could be stronger or weaker than in the pure 
gauge theory. However, as for the pure Wilson gauge theory \cite{wflow_luscher}, we find that
the probability of fields ``between the sectors'' goes to zero as 
$R_0(m)^{-10}$ also for our theory including dynamical Wilson fermions. 
By $R_0(m)$ we denote $r_0/a$ as a 
function of the quark mass, a measure for the gluonic correlation length of the system. Fig.~\ref{fig1}, left, 
shows the scaling with $a$ at fixed pion mass, but
we also verified the $R_0(m)^{-10}$ scaling as a function of 
the quark mass at fixed bare coupling.

The strong depletion of the configuration space between the sectors means that 
eventually 
the topological charge will not be properly sampled at all. 
In our case,  the algorithm 
has difficulties 
and auto-correlations are large in particular for our $a=0.048$\,fm 
ensembles. They have to be
 controlled in order
to obtain reliable MC results and errors.
We first look at the auto-correlation functions of the observables under study and, in order to quote a safe error estimate for our results, we follow the method developed in \cite{virotta_errors} to evaluate the auto-correlation times
$\texp$ and $\tint$.

\begin{figure}[t!]
\centering
\includegraphics[trim= 0 1pt 0 0, width=.6\textwidth]{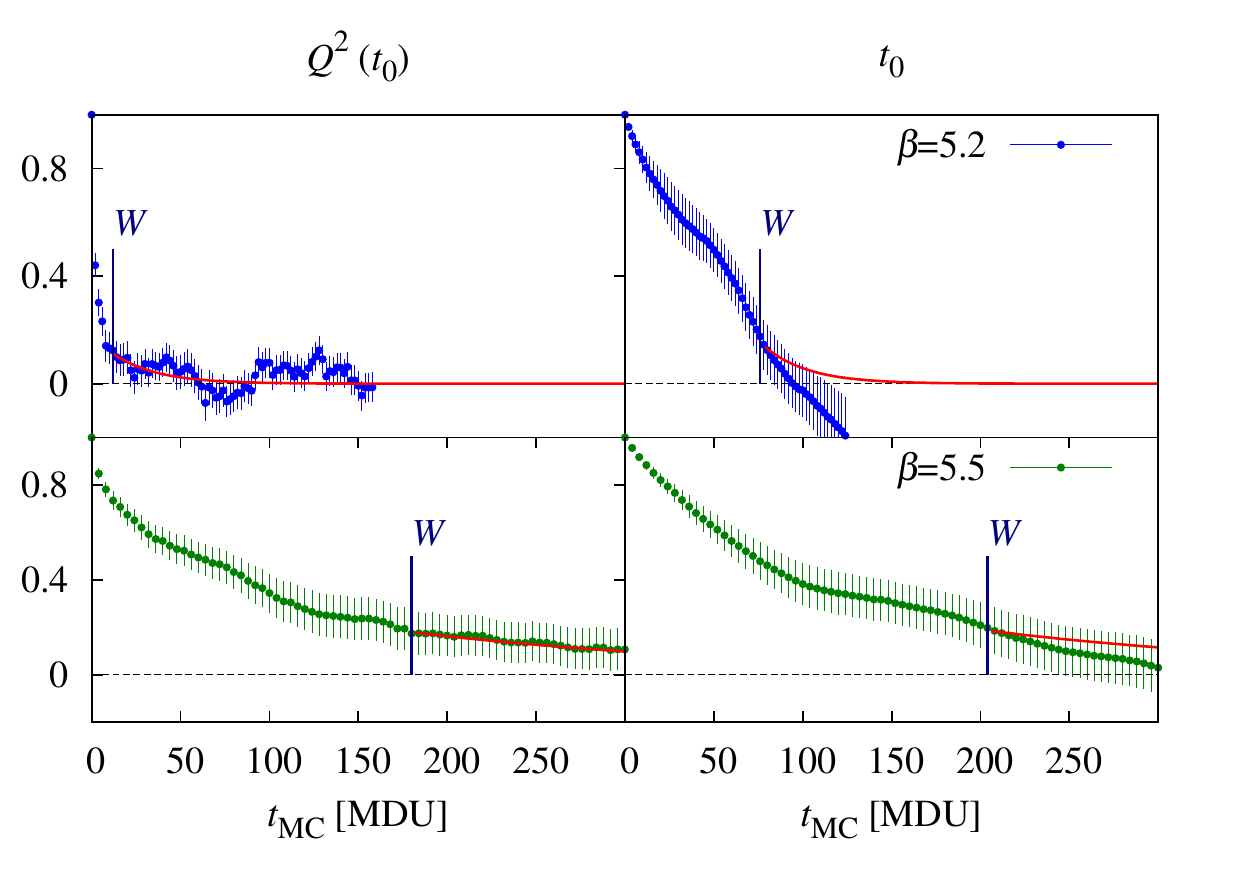}
\caption{Normalised auto-correlation functions
at $a=0.075$~fm, 
$m_\pi\approx280$~MeV (top), and at $a= 0.048$~fm,
$m_\pi=340$~MeV (bottom). The red curves are our estimates of the contribution of the tails of $\rho$.}
\label{fig:rhos_t0_Q2}
\end{figure}

In Fig.~\ref{fig:rhos_t0_Q2} we plot some examples, for two lattice spacings, 
of the auto-correlation functions $\rho(t_\mathrm{MC})$, normalised as
$\rho(0)=1$, of $t_0$ and $Q^2(t_0)$. 
In both cases auto-correlations are under reasonable control. 
For our
lattice spacings, $t_0$ shows larger auto-correlations than the topological charge. It is in fact a good estimator of the exponential auto-correlation 
time, better than the $Q^2$ used in \cite{virotta_errors}.

Recent studies \cite{openBC} showed that when open boundary conditions (along the Euclidean time) are employed, the MC sampling of topology is significantly
accelerated.  We observe that, at our largest lattice spacing, 
the ALPHA Collaboration open BC simulation has similar 
auto-correlations as the periodic BC one, as shown in Fig.~\ref{fig1}, right. 
Note that the two simulations do not use exactly the same algorithm.

\begin{figure}[t!]
\centering
\includegraphics[width=.4\textwidth]{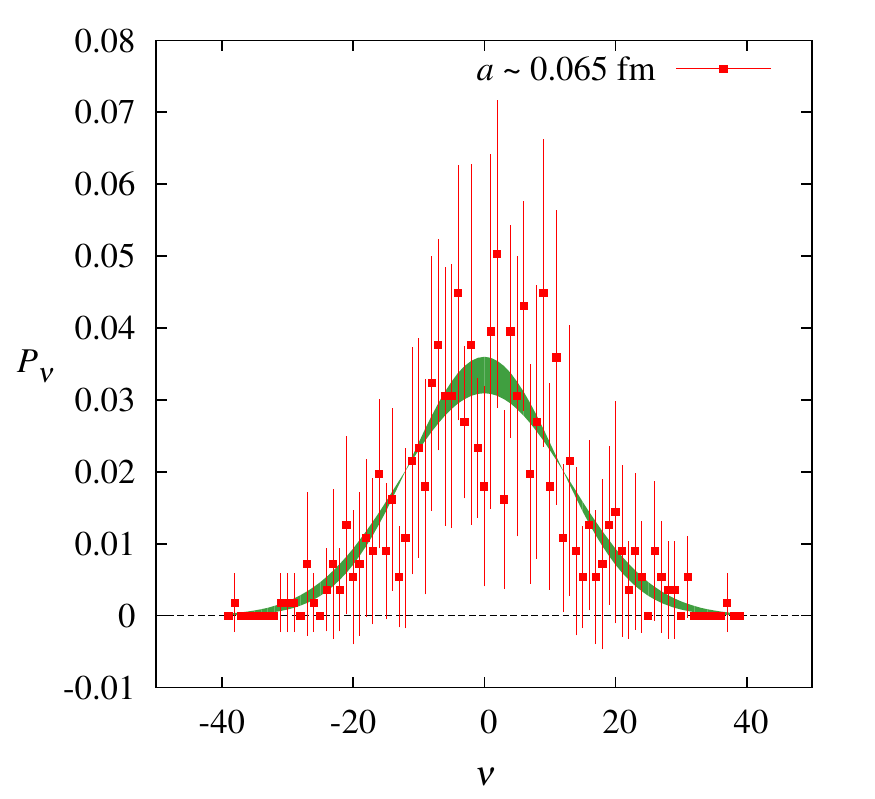}
\includegraphics[width=.55\textwidth]{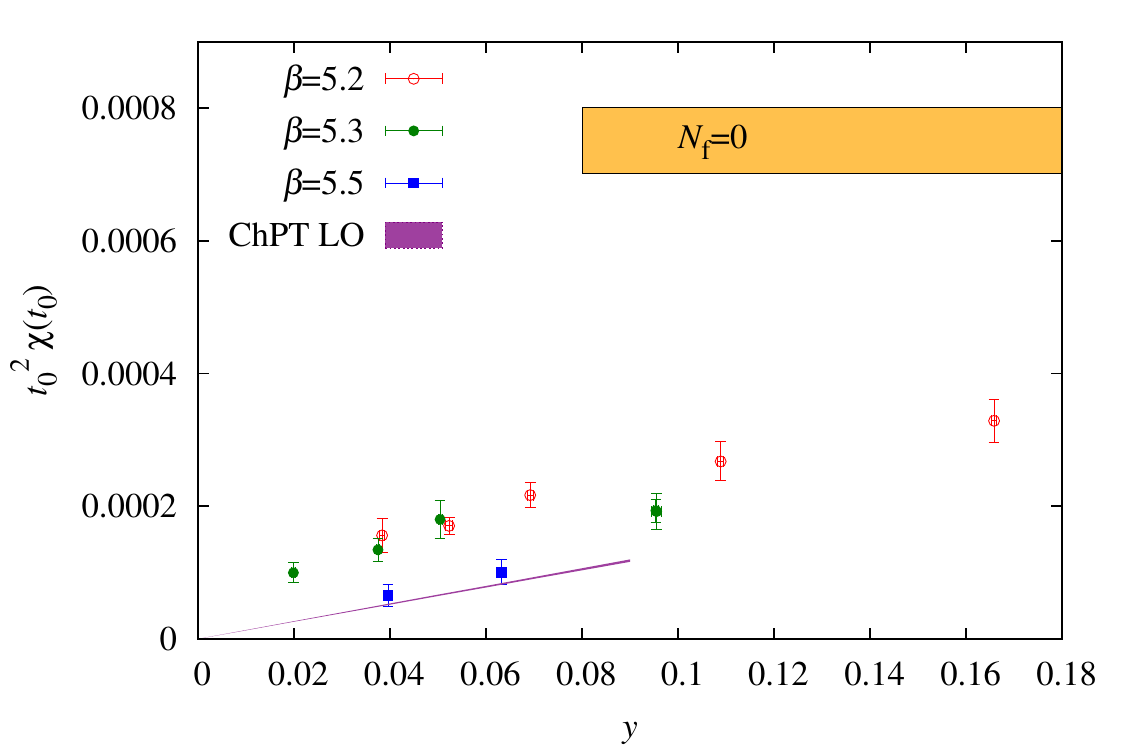}
\caption{{\it Left:} probability distribution of the topological charge at $t=t_0$ of our intermediate lattice spacing. For this ensemble 
$m_\pi=192$ MeV and $t_\mathrm{MC} = 1100$~MDU. {\it Right:} the susceptibility in units of $t_0$; the large mass asymptotic value
from $\Nf=0$ is indicated as a
horizontal error band, while the lowest order
chiral perturbation theory prediction 
(using our results for $t_0f_\pi^2$) is the purple band. 
}
\label{fig3}
\end{figure}

A further very useful test of the quality of the ensembles is to look at the distributions of the topological charge. We define the observable
\begin{equation}
f_\nu(Q) = \theta(Q - (\nu-\frac12)) \theta(\nu+\frac12 - Q) \,, \quad \nu \in \mathbb{N} \,,
\end{equation}
whose mean value $\langle f_\nu(Q) \rangle = P(\nu)$ is the 
probability of the topological charge to be in the unit length
interval around $\nu$. It is a standard observable whose error
can be computed as above. In large volume $P(\nu)$ approaches 
a Gaussian \cite{qtop_dist}. 
We find this well realised when the Monte Carlo history is at least $\approx 20 \texp$,
 (e.g. Fig.~\ref{fig3}, left
 ). Our errors render corrections to the gaussian behaviour invisible.

We now restrict ourselves to the ensembles with a length of
at least $\approx 20 \texp$ and investigate the quark mass dependence of
the susceptibility $\chi(t_0)=\langle Q^2(t_0) \rangle/(L^3\cdot T)$.
From the comparison to the $\Nf=0$ result \cite{nf0_susceptibility,wflow_luscher} in Fig.~\ref{fig3}, right, 
 the strong suppression of the susceptibility caused by the sea quarks is evident. 
However, also lattice spacing effects are clearly visible.
Note that $\chi$ has dimension (mass)$^{-4}$, where scaling violations,
e.g. simply induced through the scale setting, are strongly enhanced.
Similar scaling violations have been observed in \cite{qsusc_milc}.
When the lattice spacing is reduced down to $a=0.048$ fm, we obtain a result in rough 
agreement with the leading order of the chiral expansion of the susceptibility
\begin{equation}
\chi = \frac{m}{2} \Sigma (1 + \rmO(m)) = \frac{f_\pi^2 m_\pi^2}{8} + \rmO(m_\pi^4)  \,.
\end{equation}

\section{Scales from the gradient flow}
In scale setting~\cite{lat13:rainer}, statistically precise scales 
that mildly depend on the quark masses are particularly welcome.
A scale closely related to $t_0$ is $w_0$, defined by~\cite{wflow_BMW}
$
\left. t \frac{d}{dt} [t^2 E(t)]\right|_{t=w_0^2} = 0.3 \,.
$
Despite the large auto-correlations shown above, both $t_0$ and $w_0$ 
are more precise than $r_0$~\cite{Sommer:1993ce}; their variance is 
{\em very} small.

\begin{figure}[t!]
\centering
  \includegraphics[width=.47\textwidth]{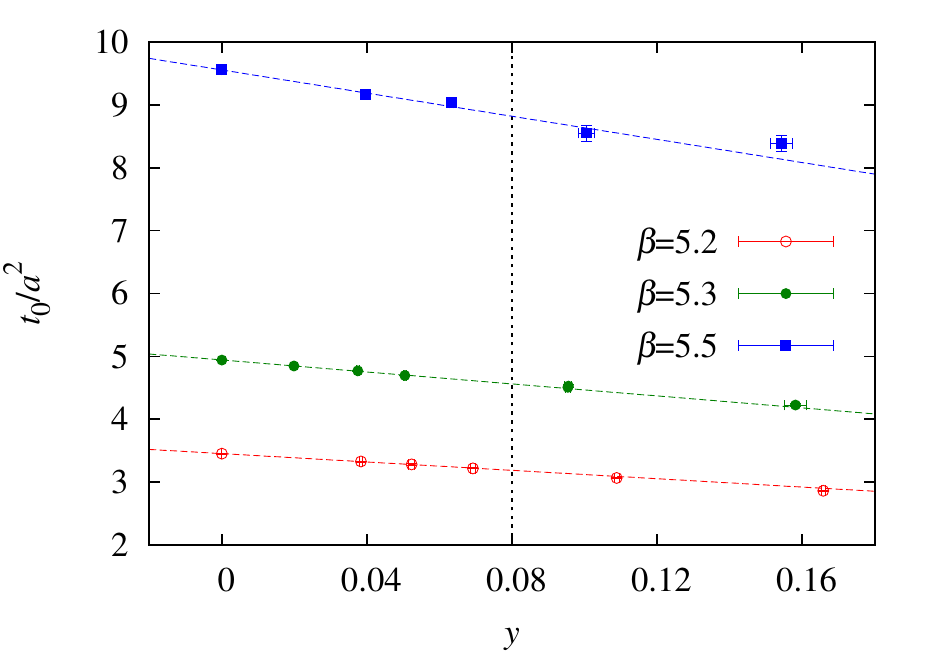}
\includegraphics[width=.47\textwidth]{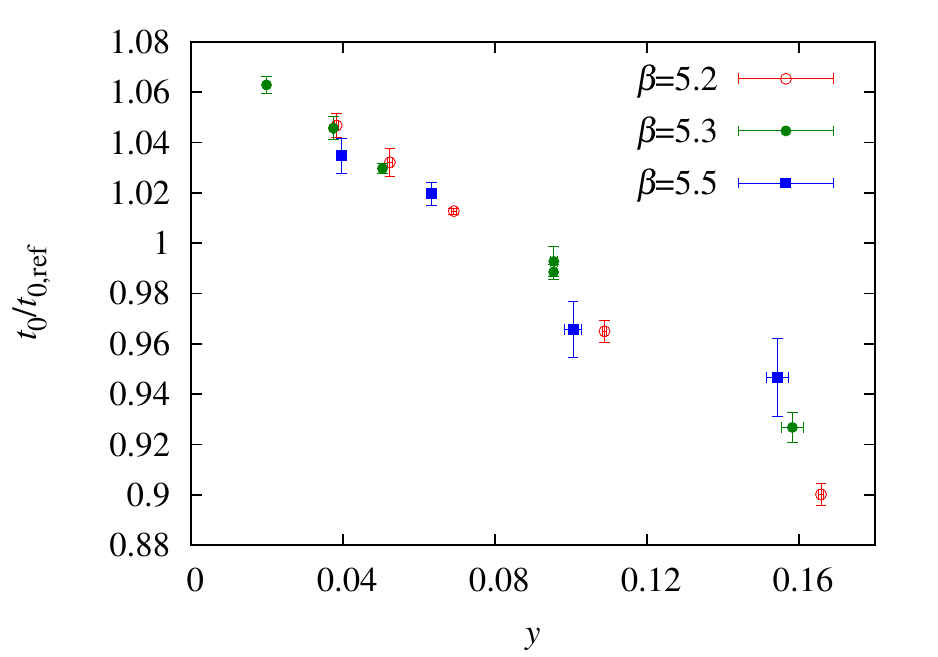}
\caption{{\it Left}: behaviour of $t_0/a^2$ as a function of the pion mass; to have control on the extrapolation we use different functions and ranges in $y$ and we quote, as final result, the linear extrapolation with $y \leq 0.08$, corresponding to $m_\pi \leq 390$ MeV. {\it Right}: for each $\beta$, $t_{0,\rm ref}$ is the value of $t_0$ interpolated to $y=0.08$.}
\label{fig4}
\end{figure}

At fixed $\beta$ a dependence on the renormalised quark mass is present
but not very strong. We linearly extrapolate the three scales using the quantity $y=m_\pi^2 t_0$, defined at finite mass, as shown in Fig.~\ref{fig4}, left. 
In the future we will incorporate the asymptotic behaviour of
chiral perturbation theory into the extrapolation of
$t_0$~\cite{Luscher:2013vga}.
An interesting question is whether there are mass-dependent 
cutoff effects. None of these are visible in ratios 
such as $t_0/t_{0,\rm ref}$ (Fig.~\ref{fig4}, right). 
Within our good precision this ratio is described by a universal curve.
At this point we note that in our $\rmO(a)$ improved
action we have neglected 
a small term $a\bg \mq \tr F_{\mu\nu}F_{\mu\nu}$~\cite{impr:pap1}. 
The reason to neglect it was 
that both $\bg$ is very small at 1-loop
\cite{pert:1loop}
and the bare subtracted quark masses $a \mq$ are very small.
We can now verify that, with few-per-mille precision, 
no residual $\rmO(a)$ effects are present in $t_0$. 
All statements made hold also for $w_0$ and $r_0$,
apart from a worse precision for the latter.

To convert our results to physical units we use the lattice spacings
of \cite{alpha:lambdanf2}, based on $\fK=155$ MeV.
We then extrapolate linearly in  $a^2/t_0$ to the
continuum limit, finding:
\begin{align}
    \tnotc &= 0.02396(37) \,{\rm fm}^2  \,, \quad 
    \wnotc = 0.1776(13) \,{\rm fm} \,,
    \label{eq:t0_chiral}
    \\
    \tnotp &= 0.02356(36) \,{\rm fm}^2  \,, \quad 
    \wnotp = 0.1757(13) \,{\rm fm} \,,
    \label{eq:t0_phys}
\end{align}
where ``phys'' indicates the physical point, given by 
the physical pion mass (and $\fK$).
A comparison to other determinations of these scales in physical units 
needs care, since it depends on how that scale was set originally, 
e.g. in \cite{wflow_BMW} the mass of the $\Omega$ baryon was used. A proper discussion of the dependence on the number of flavours requires to
consider specific ratios. We now turn to those.

\section{Dynamical quark effects}
\begin{figure}[t!]
\subfigure[$t_0/r_0^2$]{
\centering
\includegraphics[width=0.3\textwidth]{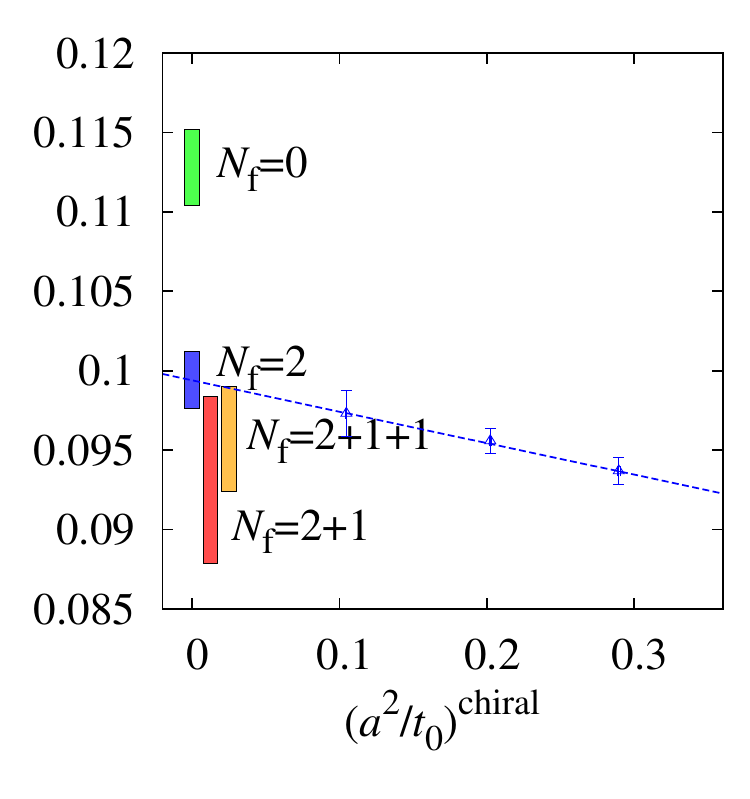}
}
\subfigure[$t_0/w_0^2$]{
\centering
\includegraphics[width=0.3\textwidth]{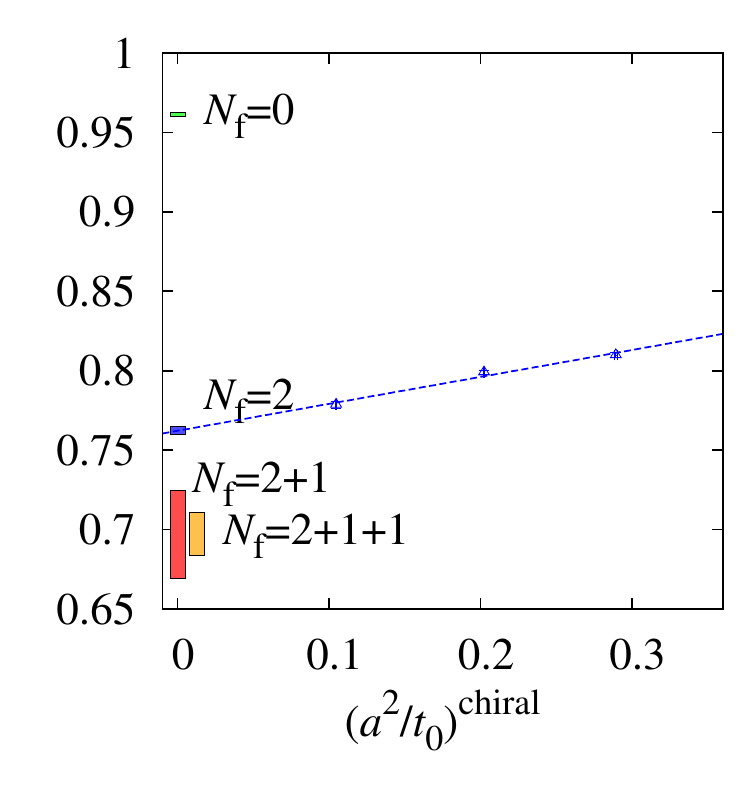}
}
\subfigure[$w_0^2/r_0^2$]{
\centering
\includegraphics[width=0.3\textwidth]{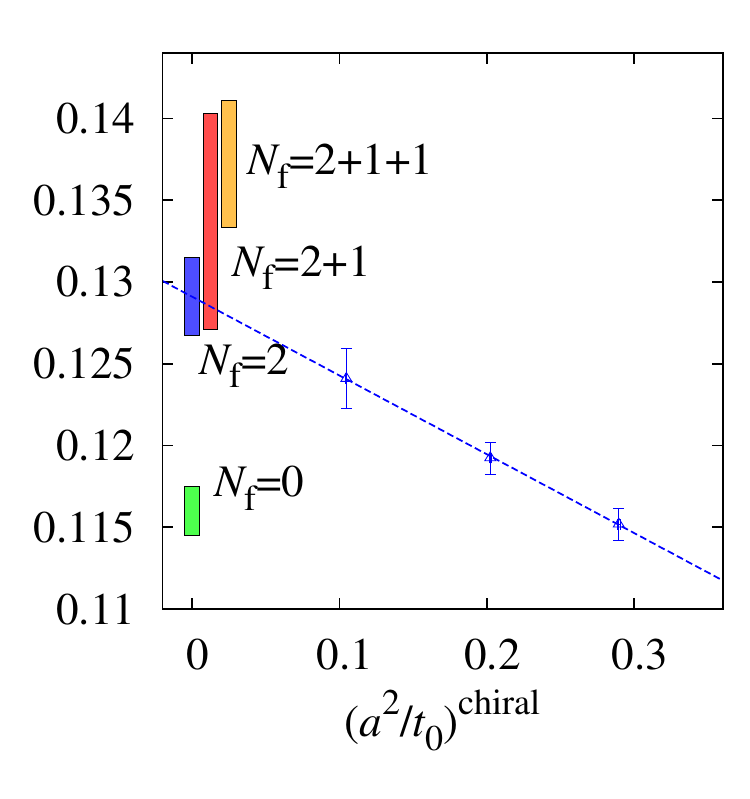}
}
\caption{Continuum extrapolation and flavour number dependence of
ratios of scales.
}
\label{fig:ratios}
\end{figure}

The three possible ratios obtained by combining $t_0/a^2$, $(r_0/a)^2$ and $(w_0/a)^2$ 
are extrapolated to the physical point as discussed above. 
We then 
approach the limit $a\rightarrow0$ of all three ratios by a linear
extrapolation in $a^2/t_0$, shown in Fig.~\ref{fig:ratios}.
Here  $t_0/r_0^2$ has the smallest discretisation effects.
The Figure also shows a 
comparison with results with a different number of flavours. 
Those for $\nf=0$ either come directly from \cite{wflow_luscher} or from our analysis of $\langle E(t)\rangle$ 
computed there. 
For $\nf=2+1$, we estimated $t_0/r_0^2$ and $w_0^2/r_0^2$ by combining $r_0=0.480(11)$~fm \cite{r0_RBC} with 
$\sqrt{t_0}=0.1465(25)$~fm and $w_0=0.1755(18)$~fm~\cite{wflow_BMW}. 
Also $t_0/w_0^2$ is computed from those numbers neglecting error correlations, which most certainly
yields a strong overestimate of the error. Finally for
$\Nf=2+1+1$ we combine 
$r_0/r_1=1.508(5)$~\cite{Bazavov:2011nk} (neglecting a difference between $\Nf=2+1$ and $\Nf=2+1+1$), $\sqrt{t_0}/w_0=0.835(8)$ and 
$r_1/w_0=1.790(25)$~\cite{wflow_nf211}. 

The ratios shown in Fig.~\ref{fig:ratios} demonstrate that the 
$\Nf=0$ and the $\Nf=2$ theories differ quite strongly for these
purely gluonic observables. This is interesting since we are looking at infrared-dominated quantities -- non-perturbative features of the theory. 
The effects of the heavier quarks,
strange and charm, appear to be less pronounced, but still noticeable.  
Of course, for a very heavy quark, decoupling is expected in the sense
that such dimensionless low energy quantities should agree
for theories with $\nf$ and $\nf-1$ quarks up to corrections 
suppressed by inverse powers of the mass $m_{\nf}$ 
of the heaviest quark.

\section{Conclusions}
In this work we have studied the topological charge and the scales $t_0$ and $w_0$ for $\Nf=2$ $\rmO(a)$-improved Wilson fermions. We demonstrated the quality
of our ensembles via empirical tests such as the distribution of the topological charge and its susceptibility. Both turn out to be in agreement with 
theoretical expectations, even if the latter is affected by large discretisation effects. We verified that when the statistics is at least $20 \texp$ auto-correlations are under reasonable control and error estimates are possible down to lattice spacings $\approx$ 0.05 fm. 
The dynamical separation of the topological
 sectors in the $\Nf=2$ theory is very similar to the pure gauge theory. 
 
The expected suppression of topology by dynamical fermions is observed.
We investigated the $\Nf$ dependence of $t_0$, $w_0$ and $r_0$
by a comparison to data in the literature.
The use of the full CLS ensembles and the mild dependence on the quark mass 
allowed for a controlled extrapolation to the two-flavour continuum theory. 
Ratios of the scales show a rather significant effect of the two light 
dynamical 
fermions, but, interestingly, already the effect of the heavier strange
quark is not as pronounced. 

\acknowledgments
We would like to thank M. L\"uscher for sharing the $\Nf=0$ data, M. Della Morte, T. Korzec and S. Schaefer for the configurations of the openBC ensemble. 
We had access to HPC resources in the form of
a regular GCS/NIC project$^1$, a JUROPA/NIC project\footnote{
http://www.fz-juelich.de/ias/jsc/EN/Expertise/Supercomputers/ComputingTime/Acknowledgements.html}
and through PRACE-2IP, receiving funding from the
European Community's Seventh Framework Programme (FP7/2007-2013) under grant
agreement RI-283493.
This work is supported in part by the grants SFB/TR9 of the Deutsche Forschungsgemeinschaft. 

\bibliography{PoS2013} 
\bibliographystyle{JHEP}


\end{document}